# Dielectric response of a ferroelectric nematic liquid crystalline phase in thin cells

Nataša Vaupotič,[1,2] Damian Pociecha,[3,*] Paulina Rybak,[3] Joanna Matraszek,[3] Mojca Čepič,[2] Joanna M. Wolska,[3] and Ewa Gorecka[3]

[1] Department of Physics, Faculty of Natural Sciences and Mathematics, University of Maribor, Koroška 160, 2000 Maribor, Slovenia
[2] Jozef Stefan Institute, Jamova 39, 1000 Ljubljana, Slovenia
[3] Faculty of Chemistry, University of Warsaw, Żwirki i Wigury 101, 02-089 Warsaw, Poland
[4] Faculty of Education, University of Ljubljana, Kardeljeva ploščad 16, 1000 Ljubljana, Slovenia

**Abstract**

We studied dielectric properties of a polar nematic phase ($N_F$) sandwiched between two gold or ITO electrodes, serving as a cell surfaces. In bulk, $N_F$ is expected to exhibit a Goldstone mode (phason), because polarization can uniformly rotate with no energy cost. However, because the coupling between the direction of nematic director and polarization is finite, and the confinement, even in the absence of the aligning surface layer, induces some energy cost for a reorientation of polarization, the phason dielectric relaxation frequency is measured in a kHz regime. The phason mode is easily quenched by a bias electric field, which enables fluctuations in the magnitude of polarization to be followed in both, the ferronematic and nematic phases. This amplitude (soft) mode is also influenced by boundary conditions. A theory describing the phase and amplitude fluctuations in the $N_F$ phase shows that the free energy of the system and, consequently, the dielectric response are dominated by polarization-related terms with the flexoelectricity being relevant only at a very weak surface anchoring. Contributions due to the nematic elastic terms are always negligible. The model relates the observed low frequency mode to the director fluctuations weakly coupled to polarization fluctuations.

**Introduction**

The discovery of a ferroelectric fluid (ferronematic phase, $N_F$) few years ago immediately caught a lot of attention [1-7]; the spontaneous electric polarization in the $N_F$ phase is comparable to that of solid ferroelectrics and is of the order of $10^{-6}$ C/cm$^2$ [8], which, combined with fluidity, makes the ferroelectric nematics potentially very attractive for future applications. In the $N_F$ phase, the longitudinal molecular dipole moments align in the same direction, which breaks the up/down symmetry of the average direction of the long molecular axis (given by a unit vector - director). The $N_F$ phase is a very interesting topic also for fundamental studies, because a full understanding of the ferroelectric order in soft materials is still at an early stage. Namely, for decades it was believed that the dipole order in soft matter is a secondary effect induced by steric interactions and requires at least some degree of a positional order [9]. For a fluid phase having a high electric polarization one can expect a giant low-frequency dielectric permittivity due to two dielectrically active relaxation modes, which can be regarded as a Goldstone-like (phason) and Higgs-like (amplitude) modes, the modes that are inherent to the Mexican-hat-form of the displacement potential [10].

**Experimental**

The dielectric permittivity was measured in 1 Hz – 10 MHz frequency ($f$) range using a Solatron 1260 impedance analyzer, which enabled application of bias field up to 40 V. The amplitude of the ac measuring field was 0.01 V/μm or less, and it was checked optically that this voltage is below the Fréedericksz transition threshold. Material was placed in 3 to 10 μm-thick glass cells with ITO or gold electrodes with no polymer alignment layer. The relaxation frequency, $f_r$, and dielectric strength, $\Delta\varepsilon$, of the mode were evaluated by fitting the complex dielectric permittivity to the Cole-Cole formula, $\varepsilon - \varepsilon_\infty = \sum \frac{\Delta\varepsilon}{\left(1+\frac{if}{f_r}\right)^{1-\alpha}} + i\frac{\sigma}{2\pi\varepsilon_o f}$, where $\varepsilon_\infty$ is a high frequency dielectric constant, α is a distribution parameter of the mode and σ is a low frequency conductivity.

**Results and Discussion**

To explore collective motions of molecules in the ferroelectric nematic phase, the dielectric dispersion measurements were performed in a broad temperature range for homologues of a model ferronematogen, RM-734 [1], differing in the length of the lateral chain (Figure 1) [11].

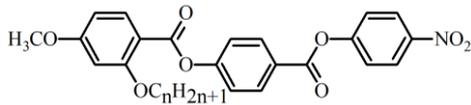

Figure 1. Molecular structure of the studied nematogens; $n = 3, 4$ and $5$.

The overall dielectric response is similar for all the studied materials (Figure 2), although the frequency range in which the dielectric modes are active decreases with an increasing lateral chain length, due to an increasing material viscosity. The dielectric response was studied in cells with gold or ITO electrodes, which were not covered with polymer layers. Such cells were chosen purposely, because cells with dielectric layers on electrodes, although commonly used to obtain a proper alignment of liquid crystalline (LC) phases and to block charge injections to the sample, cause additional effects in dielectric measurements [12]. Thin dielectric layers (polymer) act as an additional, large capacitance in a serial connection with the capacitor filled with a liquid crystal material. However, it should be noted that the capacitance of the LC layer becomes comparable to that of polymer layers when studying giant dielectric constant materials. In such a case the interpretation of dielectric measurements become complicated, as the impedance analyzer detects the apparent conductance and capacitance of the whole electric circuit and thus the dielectric constant determined from the equivalent capacitance and dielectric loss determined from equivalent conductance are strongly affected by the presence of polymer layers. Another problem may appear when the studied LC samples exhibit a low resistivity or a high electric polarization. If the polarization vector follows the applied electric field without a time delay (real Goldstone-type relaxation), the apparent resistivity of the LC slab, $R$, may be as low as 1 Ohm, so only dielectric layers are charged and their capacitance, $C_D$, is measured, yielding a relaxation mode ($PCG$-mode) with a characteristic time $\tau = RC_D$ [12]. Even when using cells with gold or ITO electrodes without dielectric layers, one has to be very careful when interpreting experimental results, because a thin layer of the studied LC material close to the surface, over which polarization will rotate from the direction preferred by the surface to the direction preferred by bulk, can cause a similar effect as surface coating. Thus, dielectric permittivity components, $\varepsilon'$ and $\varepsilon''$ on Figure 2 (and then also on Figure 3) should be understood as apparent values, which are related to the real and imaginary part of the dielectric constant of the liquid crystal in a complex way, as given in Appendix A.

In the paraelectric nematic phase a single relaxation mode is detected. The mode relaxation frequency ($f_r$) decreases and its strength ($\Delta\varepsilon$) increases critically when the temperature is lowered, which is a typical behavior for the dielectric relaxation mode originating in the collective excitation of the local electric polarization, i.e. from the excitation of the order parameter of the lower temperature ferroelectric phase (soft mode). Because the correlation length of such fluctuations grows near the phase transition temperature, the mode frequency critically decreases and its strength increases reaching a maximum at the Curie temperature, $T_c$ (Figures 2a and 2c).

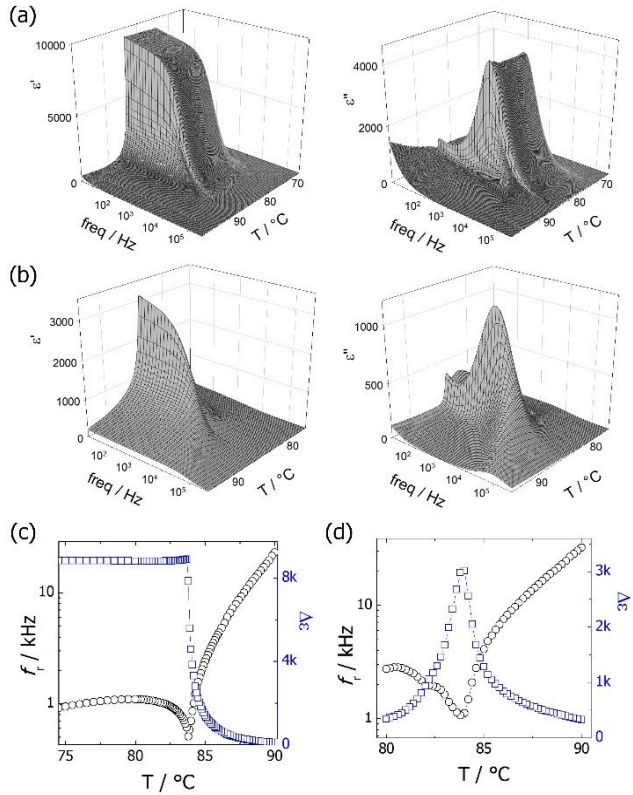

**Figure 2**. Dielectric properties of the compound with $n = 3$ measured in a 10 μm-thick cell with gold electrodes and without aligning polymer layers. The real ($\varepsilon'$) and imaginary ($\varepsilon''$) part of the dielectric permittivity vs. temperature ($T$) and frequency (freq) measured on cooling a) without a bias field and b) under applied bias field 0.4 V μm$^{-1}$. The relaxation frequency, $f_r$ (black circles), and dielectric strength, $\Delta\varepsilon$ (blue squares), of the relaxation mode c) without the bias field and d) under applied bias field, evaluated from the data presented in (a) and (b), respectively.

The dielectric response in the N$_F$ phase is dominated by a very strong mode, the origin of which can be attributed either to the director fluctuations (phason mode) as will be discussed in detail in the next section, or to the $PCG$-mode proposed in [12]. The dielectric strength of this mode, exceeding 10$^4$, is nearly temperature independent below $T_c$. A small decrease of the mode relaxation frequency near $T_c$ is caused by a contribution of the amplitude mode, because near $T_c$ it is expected that also the amplitude mode should strongly affect the dielectric response. A non-critical decrease of the mode relaxation frequency far from $T_c$ is due to a gradual increase of viscosity of the system on cooling. By applying a bias electric field, it was possible to observe that in the N$_F$ phase, close to the transition to the paraelectric phase, indeed two modes contribute to the dielectric response: a strong phason (or possibly $PCG$) mode and a much weaker amplitude mode (Figure 3). Interestingly, while the relaxation frequency of the amplitude mode slightly increases with increasing bias field, the relaxation frequency of the strong, low-frequency mode decreases, which is opposite to the typical behavior observed in the ferroelectric smectic C* phase [13], but similar to the behavior found previously in the ferroelectric smectic A (SmAP$_F$) phase [14]. By increasing the bias electric field, the low frequency mode can be completely suppressed and the critical voltage that quenches the mode increases as temperature is lowered. When the dielectric permittivity dispersion measurements are performed under a sufficiently strong bias field, the temperature evolution of the amplitude mode is clearly seen in both the N and N$_F$ phases (Fig. 2b and 2d).

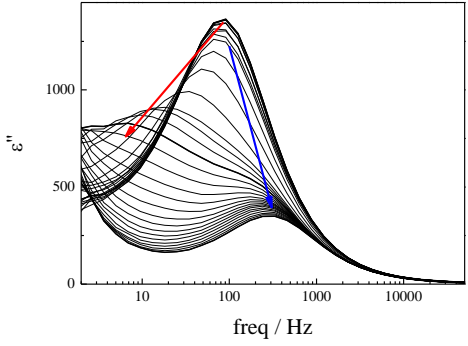

Figure 3. Frequency dispersion of the imaginary part of the dielectric permittivity ($\varepsilon''$) measured in the N$_F$ phase ($T =$ 51.7 °C, 1.5 K below $T_c$ ) for the compound with $n = 5$, under an increasing bias electric field (0 – 0.3 V/μm). Arrows show an evolution of the amplitude (blue) and phason (red) modes under a growing bias field.

As expected, the mode softens near $T_c$, i.e. its relaxation frequency decreases and its strength increases, reaching $\Delta\varepsilon = 3000$, as the system approaches the Curie temperature at both sides of $T_c$. However, the linear temperature dependence of the mode frequency holds only in the range of few degrees around $T_c$ . The steepness of the mode relaxation frequency decrease is much different at both sides of the phase transition, being 3.5, 3.7 and 6 times larger in the paraelectric N phase than in the ferroelectric N$_F$ phase for $m = 5$, $m = 4$ and $m = 3$, respectively. Far from the transition temperature at the nematic side, the mode frequency increases faster than linearly, and at the N$_F$ phase side its dependence is slower than linear. We suggest that such a nonlinear behavior is related to the temperature dependent viscosity of the material. While there is no doubt that the high frequency relaxation mode is the amplitude mode, the origin of the low frequency relaxation mode is ambiguous. As already mentioned, it could be a PCG mode or a phason mode reflecting fluctuations of the director coupled to polarization vector. For the PCG mode one can expect that its frequency will be lowered under a bias electric field, as observed experimentally, due to an increase of the apparent resistivity of LC slab [12]. However, the PCG-mode strength should grow with increasing bias field, as the apparent 'dielectric' layer, i.e. layer of LC material over which the polarization rotates to become parallel to electric field, shrinks (which increases the capacitance of the 'dielectric' layer), contrary to experimental observations, where the lower frequency mode is suppressed when the bias field increases (see Figure 3). If we make a reasonable assumption that the anchoring at the bare gold or ITO electrodes is very weak, the measured system can be considered as capacitor and resistor in parallel, without an addition of a capacitor in series and in this case the apparent $\varepsilon'$ is always the one of the liquid crystal, regardless the value of the resistivity of LC (see Appendix A). The resistivity affects the frequency dependence of $\varepsilon''$. At low resistivities no peak in the $\varepsilon''(f)$ curve is expected, just its monotonic increase with reducing frequency. At higher resistivity and high values of $\varepsilon'$ the phase mode becomes visible (Appendix A).  Thus, even though we cannot completely exclude the PCG mode, we find more arguments in favor of the phason mode and in the next section, we show that experimental observations can be explained by arguing that the lower frequency mode is a phason mode. Finally we note that at high bias voltage that aligns polarization along external field and yields uniform polarization direction across the sample thickness, the actual value of dielectric strength of the amplitude mode is measured, being as high as 3000 at the N-N$_F$ phase transition.

**Model**

The primary question is why the phase mode in the N$_F$ phase is not at zero frequency, as expected for the Goldstone mode related solely to the changes of polarization direction. In general, a non-zero frequency of the phason indicates that changes of the polarization vector direction are not free but require some energy, which could be due to a distortion of the internal structure of the phase (for example domains in ferroelectric crystals [15], helix in the SmC* phase [13] or splayed structure suggested by the Ljubljana group for the N$_F$ phase [2]); however, it can be also due to constrains imposed on the electric polarization vector by sample surfaces. When a dielectric response is studied in ITO cells (cells with transparent electrodes) a birefringent optical texture is observed in the ferroelectric nematic phase, which evidences that the director, and thus the polarization vector, is tilted with respect to the surface plane. Even for electrodes that are not covered with polymer layers, one can expect a weak director anchoring with symmetrical conditions, *i.e.* the same orientation of polarization to the surface at both cell edges, and therefore a splay deformation of polarization across the cell thickness (Fig. 4). We assume that a similar, non-uniform alignment is obtained also in cells with gold electrodes, and we tried to estimate how such constrains imposed by surfaces affect the dielectric response.

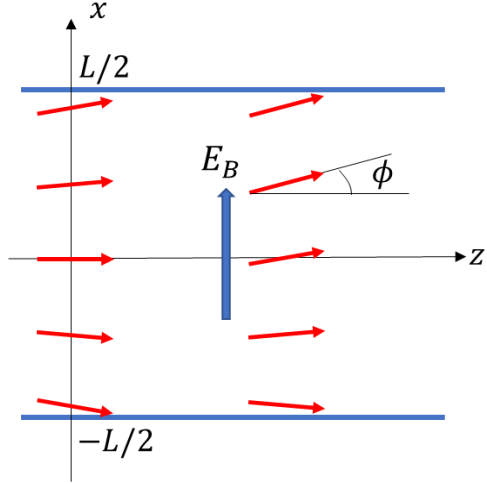

**Figure 4.** The geometry of the studied cells. The local orientation of polarization (red arrows) in the case of a weak polar homeotropic anchoring without a bias field (left) and in external bias electric field $E_B$ (right). $L$ is the cell thickness and $\phi$ the angle that defines the local direction of polarization.

To simulate the experimental conditions (*i.e.* cells with no dielectric layers on electrodes) we assumed weak polar homeotropic (HT) anchoring which favors the polarization ($\vec{P}$) at the surface to be perpendicular to the surface. As a result, there is a splay-bend deformation of polarization across the cell and, if we assume $\vec{P} \approx P\vec{n}$, also a splay-bend deformation of the director ($\vec{n}$).

To study fluctuations in the direction of $\vec{n}$ and $\vec{P}$ and in magnitude of $\vec{P}$, we use a continuous phenomenological model. In general, the free energy density ($f$) of the ferronematic LC is expressed as a sum of the elastic nematic, electrostatic and flexoelectric energy contribution, the Landau free energy density and flexodipolar terms due to a nonuniform distribution of polarization [16]. We assume that the flexodipolar terms are small and can be neglected when studying the general properties (trends) of the dielectric response of monodomain (polarization-wise) cells and express the free energy density as a sum of three contributions

$$f = f_n + f_f + f_P. \tag{1}$$

For the chosen geometry (Figure 4), the nematic free energy density ($f_n$) contains only a splay and bend deformation:

$$f_n = \frac{1}{2}K_1(\nabla \cdot \vec{n})^2 + \frac{1}{2}K_3(\vec{n} \times (\nabla \times \vec{n}))^2, \tag{2}$$

where $K_1$ and $K_3$ are the splay and bend elastic constant, respectively. The flexoelectric free energy density ($f_f$) is:

$$f_f = -\gamma_s(\vec{n} \cdot \vec{P})(\nabla \cdot \vec{n}) - \gamma_b \vec{P} \cdot ((\nabla \times \vec{n}) \times \vec{n}), \tag{3}$$

where $\gamma_s$ and $\gamma_b$ are parameters related to the splay ($e_s$) and bend ($e_b$) flexoelectric coefficient, respectively, the latter two defining the flexoelectric polarization $\vec{P}_f = e_s(\vec{n})(\nabla \cdot \vec{n}) + e_b((\nabla \times \vec{n}) \times \vec{n})$. The last contribution ($f_P$) in the free energy density is:

$$f_P = \frac{P_x^2}{2\varepsilon\varepsilon_0} - E_B P_x + \frac{1}{2}t\,(\vec{n} \cdot \vec{P})^2 + \frac{1}{4}v(\vec{n} \cdot \vec{P})^4 + \frac{1}{2}b(\nabla \vec{P})^2 \,. \tag{4}$$

The first term in eq. (4) is a self-electrostatic energy, where $P_x$ is a component of polarization along the $x$-axis and $\varepsilon$ is a dielectric constant of LC at high frequencies. This term is due to a spatial divergence of polarization, which leads to a locally induced charge $-\nabla \cdot \vec{P}$. In thin cells, polarization charge does not necessarily vanish due to the boundary effects [17], which can significantly affect the director structure in the cell [18-20]. The second term in eq. (4) is a dipole energy due to a bias DC external electric field ($E_B$) applied along the $x$-axis. The two terms in eq. (4) with $(\vec{n} \cdot \vec{P})$ are Landau terms. The parameter $t$ is temperature dependent and it is negative in the ferroelectric nematic phase. The term with parameter $v$ stabilizes a finite value of polarization. By using the term $(\vec{n} \cdot \vec{P})$ instead of simply $P$, we include a coupling between the nematic and polar order and consider the fact that the polar order does not exist without the nematic order. In Ref. [16] this coupling is included by a separate energy term. The last term in Eq. (4) describes the energy penalty for a spatial variation of polarization.

In the search for the equilibrium spatial dependence of $\vec{n}$ and $\vec{P}$, we assume that $\vec{P}$ is parallel to $\vec{n}$, i.e. $\vec{P} = P_0 \vec{n}$, where the magnitude of polarization ($P_0$) is constant across the cell. To describe the equilibrium structure in cells with weak homeotropic anchoring a planar variation in the director orientation is considered (see Figure 4):

$$\vec{n} = (\sin\phi_e(x), 0, \cos\phi_e(x)) \,, \tag{5}$$

where, for thin cells, we assume:

$$\phi_e(x) = q_s \frac{\pi}{L} x + \delta\phi_E \cos\left(q_s \frac{\pi}{L} x\right). \tag{6}$$

The anchoring strength is defined by $q_s$, which changes from $q_s = 0$ to $q_s = 1$ for no anchoring and strong anchoring cases, respectively. The first term in eq. (6) presents a spatial variation of the angle $\phi$ in the absence of an external electric field, while the second one describes a shift in $\phi$ due to a bias external electric field, with the induced rotation being the largest in the middle of the cell ($\delta\phi_E$) and reducing towards the surfaces. Experimentally used cells are not thin enough for the used harmonic ansatzes to be valid, but the chosen approach enables analytical prediction of the dielectric response of cells under external fields. To model the structure and response in thicker cells a numerical approach is required. However, a lesson learnt from the ferroelectric smectic-A phase [20], where both approaches were used, justifies analytical studies with harmonic ansatzes, as for the ferroelectric smectic-A phase it was shown that there is only a quantitative but no qualitative difference between the behavior of thin and thicker cells.

We plug the ansatzes given by eqs. (5) and (6) into the free energy density (eq. (1)) and calculate the energy per unit thickness of the cell ($F/L$), where $F = \int_{-L/2}^{L/2} f \, dx$. The energy is minimized over $P_0$ and $\delta\phi_E$. By assuming that a very weak external bias field does not affect the magnitude $P_0$ and that we are far enough from the phase transition temperature such that the flexoelectric effect also does not affect the magnitude of polarization, we find up to the second order in $q_s$

$$vP_0^2 = -t - \frac{\pi^2 q_s^2}{12\varepsilon\varepsilon_0} - \frac{b\pi^2 q_s^2}{L^2} \tag{7}$$

and

$$\delta\phi_E = \tilde{E}_B \left[1 + \frac{\pi^2}{6}q_s^2 - \pi^2 q_s^2 (1-\kappa)\frac{\xi_K^2}{L^2} - \pi q_s \frac{\xi_{\gamma s}}{L}\right], \tag{8}$$

where $\xi_{\gamma s} = \gamma_s \varepsilon \varepsilon_0 / P_0$, $\xi_K^2 = K_3 \varepsilon \varepsilon_0 / P_0^2$, $\kappa = K_1/K_3$ and $\tilde{E}_B = E_B \varepsilon \varepsilon_0 / P_0$. Assuming $\varepsilon \sim 5$ and $P_0 \sim 10^{-2}$ C/m$^2$ and $K_3 \sim 10^{-11}$N, we obtain $\xi_K \sim 10^{-9}$ m, thus the corresponding term in eq. (8) is negligible, because experimentally $L \sim 1$ μm or larger. The coefficients $\gamma_s$ and $\gamma_b$ are related to the flexoelectric coefficients as $\gamma_{s,b} \sim e_{s,b}/\varepsilon\varepsilon_0$, so the characteristic length $\xi_{\gamma s} \sim e_s/P_0$. Assuming that the flexoelectric coefficients are of the order of 1 pC/m, we find $\xi_{\gamma s} \sim 10^{-10}$ m and the last term in eq. (8) is negligible as well.

To study fluctuations in the direction of $\vec{n}$ and $\vec{P}$ we decouple the directions of the local polarization and director. Thus, $\vec{P} = P_0 \vec{p}$, where $\vec{p} = (\sin\phi_p(x), 0, \cos\phi_p(x))$ is the unit vector, and the angle $\phi_p(x)$ is

$$\phi_p(x) = \phi_e(x) + \delta\phi_p(x) e^{-i\omega\tau}, \tag{9}$$

where $\delta\phi_p(x)$ is a spatially dependent amplitude of fluctuation in the direction of polarization, $\omega$ is the angular frequency of the external field that forces fluctuation, and $\tau$ is time. The angle defining the director orientation is

$$\phi_n(x) = \phi_e(x) + \delta\phi_n(x) e^{-i\omega\tau}, \tag{10}$$

where $\delta\phi_n(x)$ is a spatially dependent amplitude of fluctuation in the direction of $\vec{n}$. To simplify the analytical results, we assume that the spatial variations of the fluctuation amplitudes are the same as the spatial variation of $\phi_e(x)$: $\delta\phi_p = \delta\phi_{p0} \cos(q_s \pi x/L)$ and $\delta\phi_n = \delta\phi_{n0} \cos(q_s \pi x/L)$. To calculate the free energy per unit cell thickness, the free energy density is Taylor expanded up to the quadratic terms in $\delta\phi_E$, $\delta\phi_{p0}$ and $\delta\phi_{n0}$. The obtained expression contains a coupling term $\delta\phi_{n0}\delta\phi_{p0}$, thus the eigenfrequencies $\omega_1$ and $\omega_2$ are proportional to the eigenvalues of the matrix

$$M = \begin{pmatrix} \dfrac{\partial^2 F}{\partial(\delta\phi_{n0})^2} & \dfrac{1}{2}\dfrac{\partial^2 F}{\partial\delta\phi_{n0}\partial\delta\phi_{p0}} \\ \dfrac{1}{2}\dfrac{\partial^2 F}{\partial\delta\phi_{n0}\partial\delta\phi_{p0}} & \dfrac{\partial^2 F}{\partial(\delta\phi_{p0})^2} \end{pmatrix}, \tag{11}$$

the proportionality factor being a viscosity, relevant for the type of fluctuation. For a weak surface anchoring ($q_s \ll 1$), we Taylor expand the eigenvalues up to the second order in $q_s$ and the first order in $\xi_b^2/L^2$, $\xi_K^2/L^2$ and $\xi_\gamma/L$, where $\xi_\gamma = (\gamma_s + \gamma_B)\varepsilon\varepsilon_0/P_0$, to find

$$\omega_1 \propto \pi^2 q_s^2 \frac{P_0^2}{\varepsilon\varepsilon_0}\left[\frac{1}{12} + \frac{\xi_b^2}{L^2} + (1-\kappa)\frac{\xi_K^2}{L^2} + \frac{2}{\pi q_s}\frac{\xi_\gamma}{L} - \tilde{E}_B^2\left(\frac{1}{\pi q_s}\frac{\xi_\gamma}{L} + (1-\kappa)\frac{\xi_K^2}{L^2}\right)\right] \tag{12}$$

and

$$\omega_2 \propto \frac{P_0^2}{\varepsilon\varepsilon_0}\left[1 - \frac{\pi^2}{6}q_s^2 + \pi^2 q_s^2 \frac{\xi_b^2}{L^2} + \tilde{E}_B^2\left(-1 + \frac{5\pi q_s}{2}\frac{\xi_\gamma}{L} + 3\pi^2 q_s^2(1-\kappa)\frac{\xi_K^2}{L^2}\right)\right]. \tag{13}$$

By comparing eqs. (12) and (13), we observe that the eigenfrequency $\omega_1$ is much lower that $\omega_2$. Experimentally observed frequencies are of the order of $100 - 1000$ Hz in zero bias field, thus we conclude that the lower frequency mode is observed. At a viscosity being of the order of 0.1 Pas, the anchoring should indeed be very weak ($q_s \sim 0.01$) to reduce the frequency to the sub kHz regime. The higher frequency mode in the $N_F$ phase, having the electric polarization of the order of $10^{-2}$ Cm$^{-2}$, is expected to be in the MHz regime.

To estimate the contribution of fluctuations in the direction of $\vec{n}$ and $\vec{P}$ to the lower and higher frequency mode, we calculate the eigenvectors corresponding to the eigenvalues and find

$$\left(\frac{\delta\phi_{n0}}{\delta\phi_{p0}}\right)_1 = \frac{24}{\pi^2 q_s^2}(1 - \tilde{E}_B^2) \tag{14}$$

for the lower frequency mode and

$$\left(\frac{\delta\phi_{n0}}{\delta\phi_{p0}}\right)_2 = -\frac{\pi^2 q_s^2}{24}(1 + \tilde{E}_B^2) \tag{15}$$

for the higher frequency mode. Keeping in mind that $q_s \ll 1$, we see that the lower frequency mode is dominated by the director orientation fluctuations (eq. (14)) and the higher frequency mode by the fluctuations in the direction of polarization (eq. (15)).

From eq. (12) we see that $\omega_1$ goes to 0, if anchoring goes to zero. Only the first two terms in eq. (12) are relevant for the calculation of the eigenfrequency in a zero bias field, because the terms with $\xi_K/L$ and $\xi_\gamma/L$ are negligibly small. We have no information on the value of the characteristic length $\xi_b$, but we estimate that $\xi_b^2/L^2 < 0.1$, because no effect of the cell thickness on the mode frequency was observed for measurements performed in cells of different thicknesses.

We must point out that the expressions given in eqs. (12) and (13) equal the diagonal terms in the matrix given by eq. (11), which means that the off-diagonal terms due to the coupling of the director and polarization fluctuation are negligible if anchoring is very weak. The coupling becomes important at higher anchoring strengths as shown in Appendix B. With increasing anchoring strength, the effect of the nematic elasticity and flexoelectricity on the mode frequency in an external bias field reduces while the effect of coupling of $\vec{n}$ and $\vec{P}$ and spatial variation of polarization gains the importance. At a very weak anchoring the major contribution to the dependence of the lower frequency on the bias field comes from the flexoelectric effect (see eq. (12)). Because a reduction in the frequency is observed experimentally (Figure 3), we predict that $e_s + e_b > 0$ for the studied material, which agrees with the wedge shape of the molecules forming the $N_F$ phase. At higher anchoring strength $\omega_1$ also reduces by increasing $E_B$; however, the effect is predominantly due to the coupling between $\vec{n}$ and $\vec{P}$ (see Appendix B).

To calculate the frequency of the fluctuations in the magnitude of polarization, we assume that the external bias field is so high that the direction fluctuations are quenched and that both the director and polarization in the equilibrium structure are along the $x$-axis. A low amplitude ac field enforces oscillations in the magnitude of polarization:

$$P(x) = P_0 + \delta P \cos\left(\frac{q_p \pi x}{L}\right) e^{-i\omega\tau}, \tag{16}$$

where we assumed that the fluctuation in the magnitude of polarization ($\delta P$) might be easier in the center of the cell than close to the surfaces. If surfaces have no effect, then $q_p = 0$ and if fluctuations close to the surfaces are not allowed, $q_p = 1$. The free energy per unit length of the cell is

$$\frac{F}{L} = -\frac{1}{4} P_0^4 \nu + \frac{1}{2}(\delta P)^2 A, \tag{17}$$

where in the lower temperature $N_F$ phase, where $t < -1/\varepsilon\varepsilon_0$ and $P_0 = \sqrt{-(t + 1/\varepsilon\varepsilon_0)/v}$, the expression for $A = A_{NF}$ is

$$A_{NF} = -\left(t + \frac{1}{\varepsilon\varepsilon_0}\right)\left(1 + \frac{\sin(\pi q_p)}{\pi q_p}\right) + \frac{b\pi^2 q_p^2}{2L^2}\left(1 - \frac{\sin(\pi q_p)}{\pi q_p}\right) \qquad (18)$$

and in the higher temperature phase, where $P_0 = 0$, and $A = A_N$, we have

$$A_N = \frac{1}{2}\left(t + \frac{1}{\varepsilon\varepsilon_0}\right)\left(1 + \frac{\sin(\pi q_p)}{\pi q_p}\right) + \frac{b\pi^2 q_p^2}{2L^2}\left(1 - \frac{\sin(\pi q_p)}{\pi q_p}\right). \qquad (19)$$

The frequency of the amplitude fluctuations is proportional to $A_{NF}$ and $A_N$. Eqs. (18) and (19) thus show that the confinement causes the frequency not to reduce to zero at the phase transition temperature defined by $t = -1/\varepsilon\varepsilon_0$. This effect is observed experimentally. On the other hand, the proposed model cannot explain the temperature dependence of the amplitude mode relaxation frequency. If the viscosity is the same in both phases, the temperature dependence of the relaxation frequency in the ferroelectric nematic phase should be twice as steep as in the nematic phase, while experimentally the steepness in the nematic phase is larger than in the ferroelectric nematic phase. Because the ratio between the two steepness coefficients strongly depends on the chosen homologue and monotonically increases if $n$ decreases, we suggest that the respective viscosity in the $N$ phase is lower than in the $N_F$ phase.

**Conclusions**

In the ferroelectric nematic phase fluctuations of polarization vector and director are coupled, and this coupling leads to two polar phason modes. In a bulk sample one of the modes should have zero relaxation frequency and the other should be in the MHz regime. However, even weak boundary anchoring in the sample with high electric polarization change this picture dramatically. The restoring force for polarization vector appears and a non-zero relaxation frequency for a reorientation of director (coupled to polarization vector) is measured, usually in the kHz regime. This low-frequency phason mode can be easily quenched when a bias electric field is applied: the mode is suppressed and shifted to a lower frequency. If the bias field is so strong that polarization is aligned along the external electric field, the amplitude mode can be followed with its critical temperature dependence. The amplitude mode is slowed down near the N-$N_F$ transition temperature, but due to the surface anchoring its frequency does not reduce to zero. Its dielectric strength at the N-$N_F$ phase transition is of order of $10^3$, thus it is reasonable to argue that measured values of dielectric constant of $N_F$ phase exceeding $10^4$ reflect the actual material properties.

**Appendix A:**

The principle of a dielectric spectroscopy is to measure the current, $I(t)$, flowing through an ideal capacitor (of capacitance $C_0$) containing a non-conducting material with a complex permittivity, $\varepsilon^* = \varepsilon' + i\varepsilon''$, when small ac voltage, $U_o e^{i\omega t}$, is applied. The part of the current that is in phase with the applied voltage gives information on the imaginary part of the dielectric constant and the part which is π/2 shifted with respect to the voltage, gives information on the real part of the dielectric constant:

$$I(t) = i\omega C U_o e^{i\omega t} = i\omega C_0(\varepsilon' + i\varepsilon'')U_o e^{i\omega t} = -\omega C_0 \varepsilon'' U_o e^{i\omega t} + \omega C_0 \varepsilon' U_o e^{i\omega\left(t+\frac{\pi}{2}\right)}. \qquad (A1)$$

However, measurements become complicated if the studied material is conducting (this conductance might be due to the ionic contamination of the sample or due to a reorientation of polarization in the electric field in material with a high electric polarization and very slow relaxation) or if dielectric (usually polymer) layers are present at the cell surfaces. In such a case the equivalent circuit is made of a 'LC resistor' and 'LC capacitor' in a parallel connection and a dielectric capacitor (polymer layers at electrodes) connected in serial (Fig. A1).

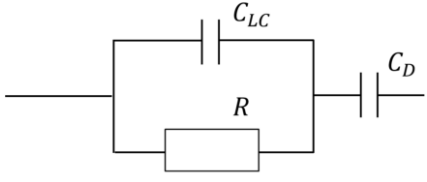

**Figure A1.** Scheme of equivalent circuit for a measuring a cell having dielectric surface layers, filled with LC material, with capacitance $C_{LC}$ and resistivity $R$. $C_D$ is a total capacitance of both surface layers.

For simplicity we consider that the dielectric layer has a very high relaxation frequency (which is often the case in experiment), thus its dielectric constant $\varepsilon_D$ is real and frequency independent. Liquid crystal has a resistivity $R$ and a complex dielectric constant with a Debye type relaxation with characteristic time $\tau_{LC}$: $\varepsilon_{LC}^* = \frac{\varepsilon_{LC}}{1+i\omega\tau_{LC}}$. The impedance of the electric circuit shown in Fig. A1 can be expressed as:

$$Z = Z_{LC} + Z_D = \frac{R}{1+iR\omega C_{LC}} + \frac{1}{i\omega C_D} \tag{A2}$$

and the current flowing through the circuit $I(t) = \frac{1}{Z}U_o e^{i\omega t}$ can be separated into a component that is in phase with the applied voltage and a component that is phase shifted by $\pi/2$ with respect to the applied voltage:

$$I(t) = \frac{R\omega^2 C_D^2(1+\omega^2\tau_{LC}^2 + R\omega^2\tau_{LC}\varepsilon_{LC}C_0)}{1+\omega^2(\tau_{LC}^2 + R^2C_D^2 + R^2\omega^2\tau_{LC}^2C_D^2 + 2R\tau_{LC}\varepsilon_{LC}C_0 + 2R^2\varepsilon_{LC}C_0C_D + R^2\omega^2\varepsilon_{LC}^2C_0^2)} U_o e^{i\omega t} +$$

$$\frac{\omega C_D(1+\omega^2\tau_{LC}^2 + 2R\omega^2\tau_{LC}\varepsilon_{LC}C_0 + R^2\omega^2\varepsilon_{LC}C_0C_D + R^2\omega^2\varepsilon_{LC}^2C_0^2)}{1+\omega^2(\tau_{LC}^2 + R^2C_D^2 + R^2\omega^2\tau_{LC}^2C_D^2 + 2R\tau_{LC}\varepsilon_{LC}C_0 + 2R^2\varepsilon_{LC}C_0C_D + R^2\omega^2\varepsilon_{LC}^2C_0^2)} U_o e^{i\omega\left(t+\frac{\pi}{2}\right)}. \tag{A3}$$

In line with eq. (A1) we now have:

$$I(t) = \omega C_0 \varepsilon_{AP}'' U_o e^{i\omega t} + \omega C_0 \varepsilon_{AP}' U_o e^{i\omega\left(t+\frac{\pi}{2}\right)}, \tag{A4}$$

where $\varepsilon_{AP}'$ and $\varepsilon_{AP}''$ are the apparent values of the real and imaginary part of a dielectric permittivity, respectively, which are related to the dielectric constant of LC in a rather complex way:

$$\varepsilon_{AP}' = \frac{1}{\omega C_0} \frac{\omega C_D(1+\omega^2\tau_{LC}^2 + 2R\omega^2\tau_{LC}\varepsilon_{LC}C_0 + R^2\omega^2\varepsilon_{LC}C_0C_D + R^2\omega^2\varepsilon_{LC}^2C_0^2)}{1+\omega^2(\tau_{LC}^2 + R^2C_D^2 + R^2\omega^2\tau_{LC}^2C_D^2 + 2R\tau_{LC}\varepsilon_{LC}C_0 + 2R^2\varepsilon_{LC}C_0C_D + R^2\omega^2\varepsilon_{LC}^2C_0^2)} \tag{A4}$$

and

$$\varepsilon_{AP}'' = \frac{1}{\omega C_0} \frac{R\omega^2 C_D^2(1+\omega^2\tau_{LC}^2 + R\omega^2\tau_{LC}\varepsilon_{LC}C_0)}{1+\omega^2(\tau_{LC}^2 + R^2C_D^2 + R^2\omega^2\tau_{LC}^2C_D^2 + 2R\tau_{LC}\varepsilon_{LC}C_0 + 2R^2\varepsilon_{LC}C_0C_D + R^2\omega^2\varepsilon_{LC}^2C_0^2)}. \tag{A5}$$

It should be noted that it is $\varepsilon_{AP}'$ and $\varepsilon_{AP}''$ that are directly determined from the experimental results.

In the limiting case of $R$ being very small one observes an 'RC mode' with a characteristic time $\tau_{RC} = RC_D$ related to a charging/discharging of the dielectric layers. In such a case no LC material properties are measured, as the capacitor filled with the LC material (cell) is not charged. In the other limiting case of $R$ being very large, LC material properties are measured but the detected dielectric constant is not directly the one of the LC. In this case two capacitors in a serial connection are measured and if the capacitance of the LC slab becomes comparable to that of dielectric layers (as can be expected for giant-permittivity materials) the equivalent capacitance will be strongly affected by the dielectric properties of polymer layers.

In a general case, both the 'RC mode' and 'LC mode' at relaxation frequencies of the LC material might be detected, which causes interpretation of measurements very complex. If $\tau_{RC}$ and $\tau_{LC}$ are comparable, the modes can be hard to distinguish, and the detected LC relaxation frequency disturbed.

Finally, let us consider the case with no surface layer, which is equivalent to making $C_D$ in eqs. (A4) and (A5) infinitely large. We find

$$\varepsilon'_{AP} = \frac{\varepsilon_{LC}}{1 + \omega^2 \tau_{LC}^2} \tag{A6}$$

and

$$\varepsilon''_{AP} = \frac{1 + \omega^2 \tau_{LC}^2 + R\omega^2 \tau_{LC} \varepsilon_{LC} C_0}{\omega C_0 R(1 + \omega^2 \tau_{LC}^2)} . \tag{A7}$$

In this case the measured $\varepsilon'_{AP}$ will always be equal to $\varepsilon_{LC}$. If $C_0 R \varepsilon_{LC} \ll \tau_{LC}$, then $\varepsilon''_{AP} = 1/(\omega R C_0)$, thus $\varepsilon''_{AP}$ will increase monotonically when $\omega$ decreases. If $C_0 R \varepsilon_{LC} > \tau_{LC}$, the mode with relaxation frequency $1/\tau_{LC}$ becomes visible.

**Appendix B:**

In this appendix we give the lower mode frequency at different strengths of a homeotropic polar anchoring. By minimizing the free energy (eq. (1)) at a general $q_s$ we find:

$$\nu P_0^2 = -t - \frac{1}{12\varepsilon\varepsilon_0} - \frac{b\pi^2 q_s^2}{L^2} + \frac{\sin(\pi q_s)}{2\pi q_s \varepsilon \varepsilon_0} \tag{B1}$$

and a very complex analytical expression for $\delta\phi_E$, so we give it only for two anchoring strengths, at $q_s = 0.25$ and $q_s = 0.5$:

$$\delta\phi_E(q_s = 0.25) = \frac{\tilde{E}_B}{0.90 + 0.03\frac{\xi_b^2}{L^2} + (0.44 - 0.41\kappa)\frac{\xi_K^2}{L^2} + 0.69\frac{\xi_{\gamma s}}{L}} , \tag{B2}$$

$$\delta\phi_E(q_s = 0.5) = \frac{\tilde{E}_B}{0.69 + 0.55\frac{\xi_b^2}{L^2} + (0.38 + 0.17\kappa)\frac{\xi_K^2}{L^2} + 0.86\frac{\xi_{\gamma s}}{L}} . \tag{B3}$$

We see that the effects of the flexoelectricity and nematic elasticity remain negligible also at higher anchoring strengths, because the terms with $\xi_K/L$ and $\xi_{\gamma s}/L$ in the denominator of eqs. (B2) and (B3) are negligible. On the other hand, the value of $\xi_b$ is not known and due to a high spontaneous polarization, it cannot be assumed that it as small as the other two characteristic lengths.

The eigenfrequencies are found by repeating the procedure given in the main paper. Again, the expressions are very lengthy even when we Taylor expand them up to the first order terms in $\xi_b^2/L^2$, $\xi_K^2/L^2$ and $\xi_\gamma/L$. For this reason, we give expressions only for the lower frequency, which is of interest to us, and only at two anchoring strengths, corresponding to $q_s = 0.25$ and $q_s = 0.5$:

$$\omega_1(q_s = 0.25) \propto \frac{P_0^2}{\varepsilon\varepsilon_0}\left[0.047 + 0.57\frac{\xi_b^2}{L^2} + (0.42 - 0.39\kappa)\frac{\xi_K^2}{L^2} + 1.4\frac{\xi_\gamma}{L}\right] -$$
$$- \frac{P_0^2}{\varepsilon\varepsilon_0}\tilde{E}_B^2\left(7.6 \cdot 10^{-4} + 0.019\frac{\xi_b^2}{L^2} + 0.74\frac{\xi_\gamma}{L} + 0.57(1-\kappa)\frac{\xi_K^2}{L^2}\right) \quad (B4)$$

and

$$\omega_1(q_s = 0.5) \propto \frac{P_0^2}{\varepsilon\varepsilon_0}\left[0.14 + 0.57\frac{\xi_b^2}{L^2} + (0.30 + 0.14\kappa)\frac{\xi_K^2}{L^2} + 1.4\frac{\xi_\gamma}{L}\right] -$$
$$- \frac{P_0^2}{\varepsilon\varepsilon_0}\tilde{E}_B^2\left(0.019 + 0.41\frac{\xi_b^2}{L^2} + 1.3\frac{\xi_{\gamma b}}{L} + 1.2\frac{\xi_{\gamma s}}{L} + (-3.59 + 3.58\kappa)\frac{\xi_K^2}{L^2}\right). \quad (B5)$$

By comparing eqs. (B4) and (B5) with eq. (12) we see that the effect of the nematic elasticity on the fluctuation frequency is negligible at all anchoring strengths, both with and without the external bias field. The flexoelectric effect cannot be neglected in the presence of the external bias field when anchoring is very weak. At higher (but still weak) anchoring strengths two additional terms appear next to $\tilde{E}_B$ in the expression for $\omega_1$, one due to the coupling between $\vec{n}$ and $\vec{P}$ and the other due to a spatial variation in the direction of $\vec{P}$. Because $\xi_\gamma/L$ is of the order of $10^{-4}$ or lower we see that the coupling term prevails already at low surface anchoring (eq. (B4)).

## Acknowledgment

The research was supported by the National Science Centre (Poland) under the grant no. 2021/43/B/ST5/00240.## References